\documentclass[aps,prb,twocolumn,showpacs,showkeys,footinbib,superscriptaddress]{revtex4}%superscriptaddress
\usepackage{amsmath}
\usepackage{amssymb}
\usepackage{graphicx}
\usepackage{bm}
\usepackage{color}
\usepackage{relsize}
\usepackage{braket}
\usepackage{textcomp}

\usepackage{hyperref}
\usepackage[all]{hypcap}

\begin{document}
\title{Phonon-limited carrier mobility in monolayer Black Phosphorous}
\author{Yuri Trushkov}
\affiliation{Skolkovo Institute of Science and Technology, 3 Nobel Street, Skolkovo, Moscow Region 143025, Russia}
\author{Vasili Perebeinos}
\email[]{v.perebeinos@skoltech.ru}
\affiliation{Skolkovo Institute of Science and Technology, 3 Nobel Street, Skolkovo, Moscow Region 143025, Russia}

\date{\today}

\begin{abstract}
We calculate an electron-phonon scattering and intrinsic transport properties of black phosphorus monolayer using tight-binding and Boltzmann treatments as a function of temperature,  carrier density, and electric field. The low-field mobility shows weak dependence on density and, at room temperature, falls in the range of 300 - 1000 cm$^2$/Vs in the armchair direction and 50 - 120 cm$^2$/Vs in the zig-zag direction with an anisotropy due to the effective mass difference. At high fields, drift velocity is linear with field up to 1 - 2  V/$\mu$m reaching values of $10^7$ cm/s in the armchair direction, unless self-heating effects are included.
\end{abstract}

%\keywords{excitons, Stark effect, transition metal dichalcogenides}
\maketitle

\section{Introduction}\label{sec:intro}

Recently, black phosphorous emerged as a new two-dimensional (2D) material, which has attracted a great deal of attention due to its excellent transport and optical properties~\cite{xia2014review,liu2014review,Ling2015bp_review,Perello2015nature,xia2014nature,engel2014photodetector,youngblood2015,Guo2016:bp_photodetect,Buscema2014bp_fet,Li2014bpfet,Chen2015bp_general,Liu2014Egbp,
Koenig2014bp_fet}. Black phosphorous  exhibits direct bandgap ranging from 0.3 eV to 1.8 eV for bulk and monolayer films respectively, filling the space between the gapless graphene and wide-gap transition metal dichalcogenides. High mobility at room temperature~\cite{Perello2015nature,xia2014nature}, strong coupling with light, and high anisotropy all contribute to huge potential in application of black phosphorous for infrared imaging~\cite{engel2014photodetector},  detection~\cite{youngblood2015,Buscema2014bp_fet,Guo2016:bp_photodetect}, and  electronic applications \cite{Li2014bpfet,Koenig2014bp_fet,Liu2014Egbp,Chen2015bp_general}. However, little is known about the interplay of the acoustic and optical phonon scattering as a a function temperature, carrier density, and electric field.

Here we calculate the electron-phonon interactions and drift velocity in a monolayer black phosphorous, within a standard tight binding approach. Our results  provide a detailed microscopic picture of phonon scattering. In particular, we find that two optical phonons with energies of around 400 cm$^{-1}$ dominate scattering at room temperatures with acoustic phonon contribution being around 30\%. The carrier density dependence of the low-field mobility is weak, such that mobilities at rooms temperature vary by a factor of three as carrier density varies by three orders of magnitude from 10$^{11}$ cm$^{-2}$ to 10$^{14}$ cm$^{-2}$. This is a consequence of a constant density of states in 2D.

\section{Theoretical Model}\label{Sec:Model}

For the transport calculations, we use Fermi Golden rule to obtain the electron-phonon scattering rates. The model ingredients include:  single-particle band structure fitted to GW calculations from Ref.~\cite{rudenko2014dft}; valence force phonon model fitted to DFT calculations from Ref.~\cite{tomanek2014phonons}; and electron-phonon coupling constants fitted to our  DFT calculations shown in Appendix~\ref{Sec:DFT}.

\begin{figure}[t]
\centering
\includegraphics*[width=10.50cm]{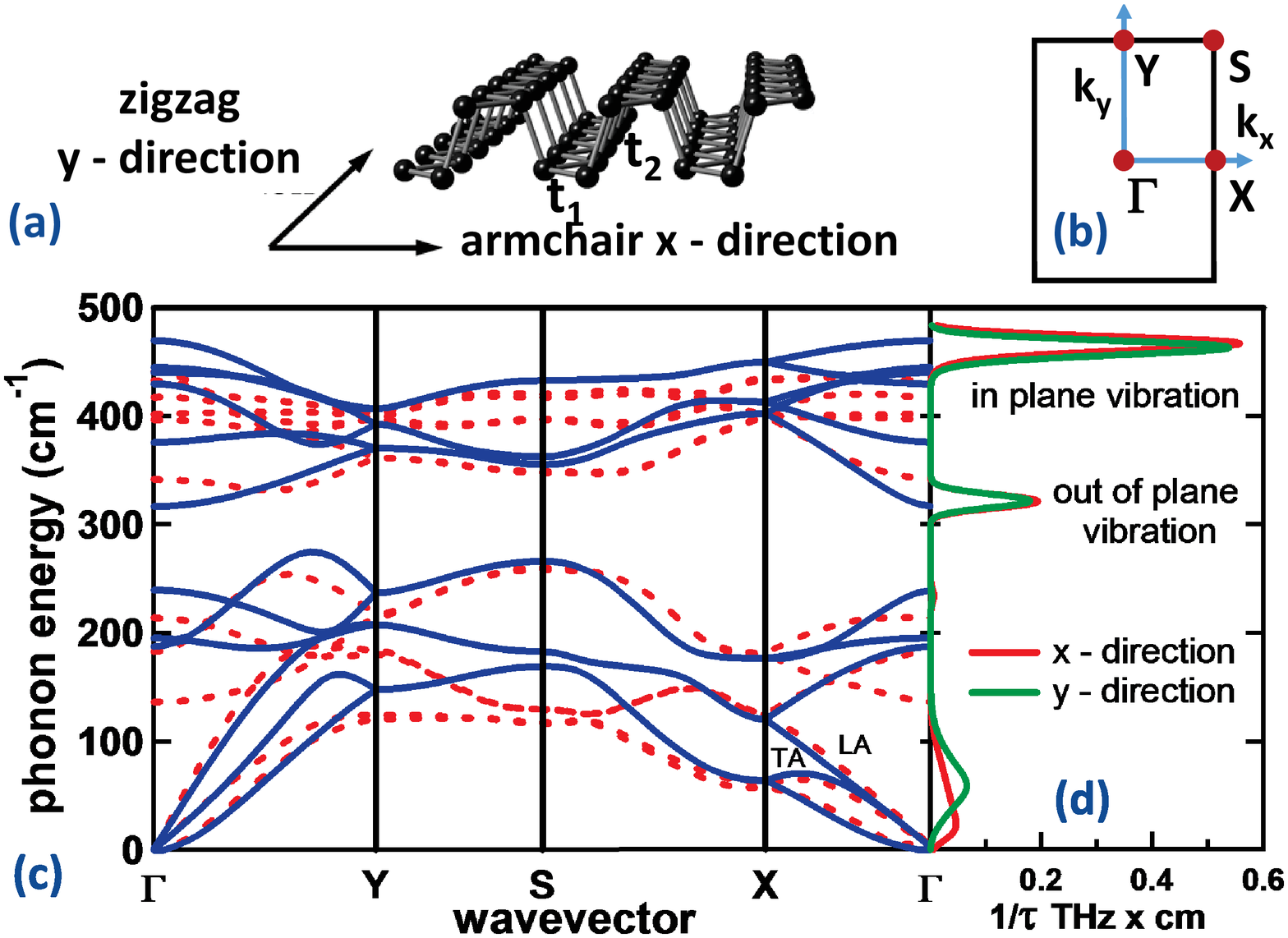}
\caption{(Color online) (a) Black phosphorus monolayer crystal structure showing tight binding hopping integrals. (b) Brillouin zone in the reciprocal space. (c) Phonon dispersion according to our model (blue solid curves) fitted to the DFT calculations in Ref.~{\protect{\cite{tomanek2014phonons}}} (red dashed curves). (d) Phonon contributions to the weighted average scattering rate in Eq.~(\ref{eq:scat}) at T=300 K and n=10$^{13}$ cm$^{-1}$ for the armchair direction (red) and zig-zag direction (blue). The two peaks at 320 cm$^{-1}$ and 465 cm$^{-1}$ correspond to the out-of-plane and in-plane vibrations near the $\Gamma$-point, correspondingly.
}\label{fig:scattering}
\end{figure}

\begin{figure}[h]
\centering
\includegraphics*[width=11.0cm]{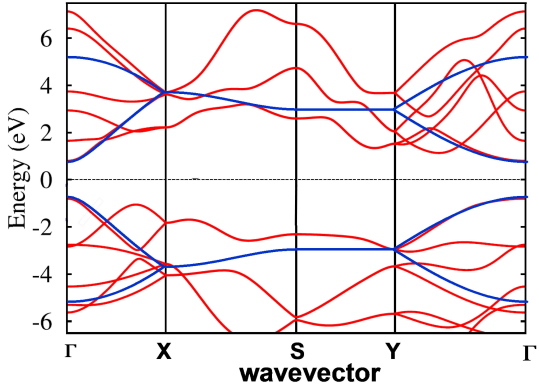}
\caption{(Color online) Bandstructure using tight binding model with $t_1=-1.15$ eV and $t_2=3.07$ eV (blue solid curves) fitted to the GW  calculations from Ref.~{\protect{\cite{rudenko2014dft}}} (red solid curves).
}\label{fig:tb_model}
\end{figure}

The single-particle band structure of black phosphorous monolayer can be obtained within the nearest-neighbor interactions in-plane $t_1=-1.15$ eV and out-of-plane $t_2=3.07$ eV hopping integrals, respectively, as shown in Fig.~\ref{fig:scattering}a. The bandgap $E_g$ and effective masses in our model are:
\begin{equation}\label{eq:tb}
\begin{split}
E_g&=2t_2-4\vert t_1\vert=1.54 \  {\rm eV} \\
m_{ex}&=m_{hx}=\hbar^2E_g/(\vert t_1\vert t_2a^2)=0.15 \ m_e \\
m_{ey}&=m_{hy}=2\hbar^2/(\vert t_1\vert b^2)=1.2 \ m_e
\end{split}
\end{equation}
where $m_e$ is the electron mass and the lattice parameters $a=4.627$ \AA \ \ and $b=3.308$ \AA \ \ are along the armchair ($x$) and the zig-zag ($y$) directions, respectively.
The bandgap is consistent with the experiment~\cite{Li2016bpgap_exp_GW} and GW calculations~\cite{rudenko2014dft,tran2014excitons,Li2016bpgap_exp_GW} and effective masses with the density functional theory calculations~\cite{Li2014Egbp_starin,Qiao2014,peng2014strain}. The overall bandstructure of our parametrization of the tight binding model agrees fairly well with that from GW calculations in Ref.~\cite{rudenko2014dft} near the top of the valence bands and bottom of the conduction bands, as shown in Fig.~\ref{fig:tb_model}.

For the phonon model, we use a two parameter Keating-type model~\cite{keating1966}:
\begin{equation} \label{eq:h_ph}
\begin{split}
H &=\sum_{<ij>}\frac{k_1}{2} \delta r_{i,j}^2 +\sum_{<ijk>}\frac{k_2}{2}\delta\theta_{i,j,k}^2 \\
&\delta \theta_{ijk}= \arccos{\frac{\vec{r_{ij}} \vec{r_{jk}}}{r_{ij} r_{jk}}} - \arccos{\frac{\vec{r_{ij}}^0 \vec{r_{jk}}^0}{r_{ij}^0 r_{jk}^0}}\\
&\delta r_{ij}=|(\vec{r_j}-\vec{r_j}^0)-(\vec{r_i}-\vec{r_i}^0)|
\end{split}
\end{equation}
where $\delta r_{ij}$ is a distance variation between the neighbouring atoms $i$ and $j$
and $\delta \theta_{ijk}$
is a variation of the angle made up by the three neighbouring atoms $i$, $j$ and $k$. Variables marked with superscript zero correspond to the equilibrium lattice structure~\cite{morita1986review,liu2014review,peng2014strain}.
The best fit to the phonon dispersion from the DFT calculations~\cite{tomanek2014phonons}, shown in Fig.~\ref{fig:scattering}c, uses the bond stretching stiffness $k_1=7.2$ eV/\AA$^{2}$ and the bending stiffness $k_2=4.4$ eV.

The electron-phonon interaction is modelled by the Su-Schrieffer-Heeger (SSH) hamiltonian~\cite{ssh1979soliton} with distance dependent hopping integrals $t_1(r_{ij})=t_{1}-g_1\delta r_{ij}$ and $t_2(r_{ij})=t_{2}-g_2\delta r_{ij}$. Using our DFT calculations of the bandgap modulation as a function of an applied in-plane and out-of-plane strain to a black phosphorus monolayer as shown in Appendix~\ref{Sec:DFT}, we find  {\it e-ph} coupling constants $g_1=1.53$ eV/\AA~ and $g_2=2.46$ eV/\AA. The Fourier transformed SSH Hamiltonian is
\begin{equation} \label{eq:h_elph2}
H_{el-ph} = \sum_{q\mu kij} M_{q\mu kij}c_{k+qj}^{\dag} c_{ki} \Big( a_{q\mu} +a_{-q\mu}^\dag \Big)
\end{equation}
where $M_{q\mu kij}$ is the {\it e-ph} coupling \footnote{Here we neglect terms corresponding to the virtual electron-hole pair excitations across the bandgap.}, $c_{k+qj}^{\dag}$($c_{ki}$) denotes creation (annihilation) of an electron in the conduction band with index $i,j=1, 2$, $a_{-q\mu}^{\dag}$ is a phonon creation operator with wavevector $-q$ and phonon band index $\mu=1 \dots 12$. Previous studies of the intrinsic mobilities used the deformation potentials to describe acoustic phonon~\cite{Qiao2014,Rudenko2016_mobility} and flexural phonon modes~\cite{Rudenko2016_mobility} scattering. Inclusion of the optical phonon scattering~\cite{Liao2015bp_mobility_abinitio} predicted lower mobilities.

The low-field electron mobility can be calculated according to Appendix~\ref{Sec:RTA}:
\begin{eqnarray}
 \label{eq:scat}
  \frac{1}{\tau_{k\alpha}} &=& \sum_{k'} S_{kk'} \frac{1-f_{0k'}}{1-f_{0k}} \Big( 1 - \frac{v_{k'\alpha}}{v_{k\alpha}} \Big) \\
  \label{eq:low_mob}
  \mu_{\alpha} & = & \frac{e\int  d^2 k
  \left(-\frac{\partial f_{0k}}{\partial E_k}\right) v_{k\alpha}^2 \tau_{k\alpha}}{\int d^2 k f_{0k}}
\end{eqnarray}
where $\alpha=x, y$ labels direction of an electric field, $S_{kk'}\propto \vert M\vert^2$ is the electron-phonon scattering rate, $f_{0k}(E_F-E_k, T)$ is the equilibrium Fermi Dirac distribution at temperature $T$ and Fermi energy $E_F$, $v_{k\alpha}$ is the band velocity at wavevector index $k$, which absorbs both wavevector direction and electron band index. The hole mobility is identical, due to the electron-hole symmetry in our model.

\section{Results}\label{Sec:results}

In Fig.~\ref{fig:scattering}d we show relative contributions of different phonons to the scattering rate in Eq.~(\ref{eq:scat}), weighted average by a distribution $\partial f_{0k}/\partial E_k$ for $n=10^{13}$ cm$^{-2}$ and $T=300$ K. The weighted average of the scattering times in $x$ and $y$ directions are very similar, as shown in Fig.~\ref{fig:scattering}d. The areas under both curves in Fig.~\ref{fig:scattering}d correspond to the scattering times $\bar{\tau}_x=63$ fs and $\bar{\tau}_y=62$ fs, which would suggest Drude mobilities to be $\mu_x=e\bar{\tau}_x/m_x=740$ cm$^2$/Vs and  $\mu_y=e\bar{\tau}_y/m_y=90$ cm$^2$/Vs, respectively. This is consistent with the full calculations of the low-field mobilities according to Eq.~(\ref{eq:low_mob}) $\mu_x=625$ cm$^2$/Vs and  $\mu_y=82$ cm$^2$/Vs.

For a high field $F$, drift velocity can be obtained from the numerical solution of the Boltzmann transport equation~\footnote{The Brillouin zone mesh of 1000 $\times$ 500
and energy cut off 0.5 eV above the bottom of the conduction band were employed to transform  Boltzmann transport equation to a matrix form
and to solve it iteratively.} for the distribution function $f_k$:
\begin{equation} \label{eq:BTE}
e F_{\alpha} \frac{\partial f_k}{\partial \hbar k_{\alpha}}=-\sum_{k'} S_{kk'}f_{k}(1-f_{k'}) - S_{k'k} f_{k'} (1-f_{k})
\end{equation}
We find that mobility from the Boltzmann transport equation solution Eq.~(\ref{eq:BTE}) differs by 5\%-10\% from the mobility calculated from Eq.~(\ref{eq:low_mob}).

\begin{figure}[t]
\centering
\includegraphics*[width=8.65cm]{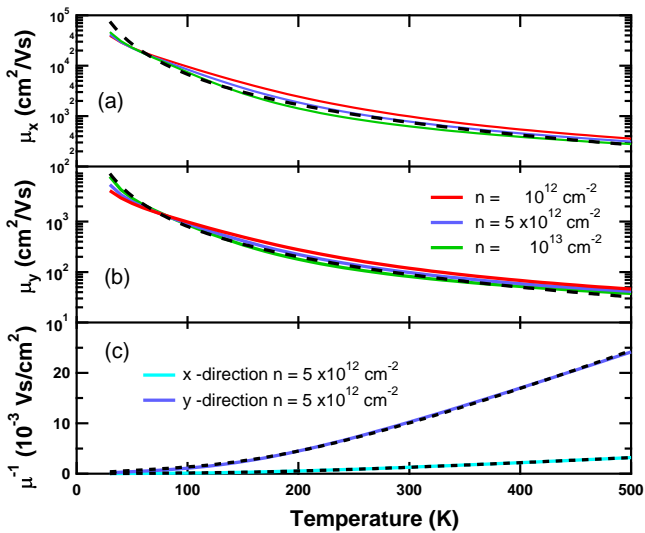}
\caption{(Color online) Calculated temperature dependence of the low-field mobility in black phosphorus along (a) the armchair $x$ and (b) the zig-zag $y$ directions at $n=10^{12}$, $5 \times 10^{12}$ and $10^{13}$  in red, blue, and  green solid curves, correspondingly. The black dashed curves show $1/T^2$ scaling to guide the eye. (c) inverse mobility at $n=5 \times 10^{12}$ cm$^{-2}$ for $x$ and $y$ directions along with the best fit to Eq.~(\protect{\ref{eq:mobfit}}), shown by the dashed curves.}\label{fig:mobility_temp}
\end{figure}

Fig.~\ref{fig:mobility_temp} shows temperature dependence of the low-field mobility for different concentrations. As expected, phonon-limited mobility decreases with temperature but exhibits several peculiarities. In the armchair direction, the room temperature mobility values are not as large as in graphene, but higher than that in  silicon and other currently employed materials confirming high potential of black phosphorus in electronics and optoelectronics. In the zig-zag direction, mobility values are about eight times lower due to the differences in the effective masses.

The overall temperature dependence appears as a power law in Fig.~\ref{fig:mobility_temp}a and \ref{fig:mobility_temp}b. However, a more detailed analysis based on the scattering rate from Fig.~\ref{fig:scattering}d suggests contributions from both acoustic and optical phonons, which can be accounted for by an empirical expression as motivated by the Matthiessen rule~\cite{Ashcroft1976book}:
\begin{equation} \label{eq:mobfit}
\mu^{-1}=\mu^{-1}_{ac}\frac{k_BT}{k_BT_{rt}}+\mu^{-1}_{op}
\frac{\exp{\left(\hbar\omega_{op}/k_BT_{rt}\right)}-1}{\exp{\left(\hbar\omega_{op}/k_BT\right)}-1}
\end{equation}
where $T_{rt}=300$ K and three fitting parameters are acoustic phonon limited mobility $\mu_{ac}$ at $T_{rt}$,  optical phonon limited mobility $\mu_{op}$ at $T_{rt}$, and optical phonon energy $\hbar\omega_{op}$. Using $\mu_{ac}=2570$ cm$^2$/Vs, $\mu_{op}=1140$ cm$^2$/Vs for $x$ direction and $\mu_{ac}=260$ cm$^2$/Vs, $\mu_{op}=160$ cm$^2$/Vs for $y$ direction, and phonon energy of  $\hbar\omega_{op}=450$ cm$^{-1}$ (or 56 meV),   Eq.~(\ref{eq:mobfit}) reproduces very well calculated mobility temperature dependence, as shown in Fig.~\ref{fig:mobility_temp}c.

\begin{figure}[t]
\centering
\includegraphics*[width=8.65cm]{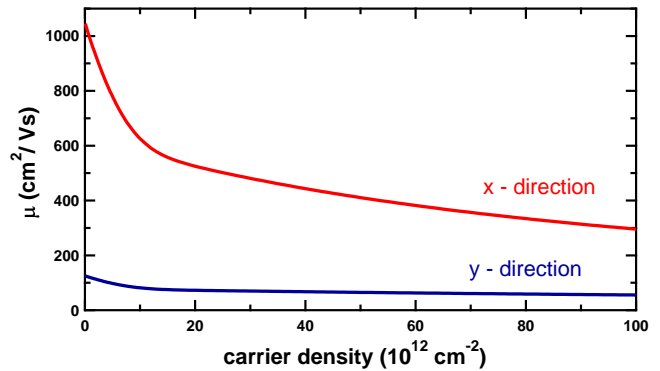}
\caption{(Color online) Density dependence of mobility at T=300 K  are shown along the armchair $x$ (red) and zig-zag $y$ (blue) directions, correspondingly. }\label{fig:mobility_density}
\end{figure}

The mobility variation with carrier density at room temperature is weak, as shown in Fig.~\ref{fig:mobility_density}. We attribute this behaviour to the constant density of states in  2D and dominant contribution of the optical phonon scattering at room temperature.
At carrier density of about $10^{13}$ cm$^{-2}$, the Fermi level coincides with the optical phonon energy. This causes suppression of the optical phonon emission due to the lack of the final density of states. As a result, mobility increases by roughly a factor of two in the limit of low carrier density. Note, that the carrier density dependence exhibits different behaviour at low temperatures below $70$ K, where acoustic phonon contribution dominates, such that mobility slightly increases as the carrier density increases.

\begin{figure}[t]
\centering
\includegraphics*[width=8.65cm]{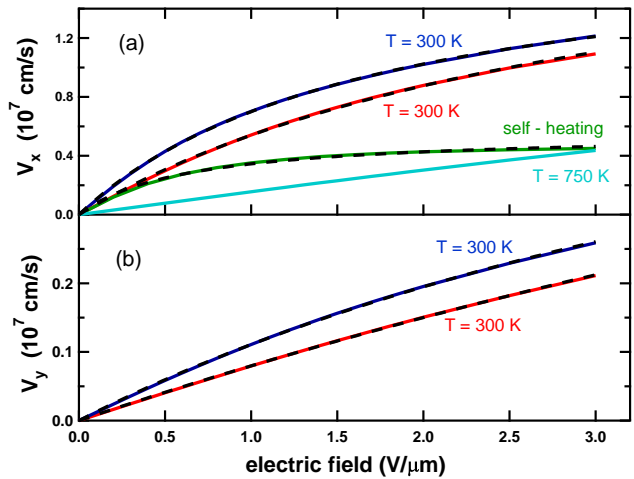}
\caption{(Color online) Field dependence of drift velocity along the armchair $x$ (a) and zig-zag $y$ (b) directions. Isothermal curves for $T=300$ K are shown for $n=10^{12}$ (red) and  $n=10^{13}$ (blue). In (a) we show isothermal curve for $T=750$ K (cyan) and calculations including self-heating effect on 300 nm SiO$_2$ at $n=10^{13}$ (green). The black dahsed curve are best fits to Eq.~(\protect{\ref{eq:vsat}}). } \label{fig:saturation}
\end{figure}

For electronic applications velocity saturation is important. At high electric fields, drift velocity for both directions is shown in Fig.~\ref{fig:saturation}a and \ref{fig:saturation}b. We find that for a constant phonon temperature, i.e. isothermal calculations, there is a weak sign of the velocity saturation due to the optical phonon scattering at the experimentally accessible field strengths. This is opposite to the case of graphene, where high energy optical phonon scattering is much stronger then the acoustic phonon scattering which leads to the velocity saturation \cite{meric2008saturation,Perebeinos2010:PRB}.
Dashed curves in  Fig.~\ref{fig:saturation} are fits to the velocity saturation expression:
\begin{equation} \label{eq:vsat}
V=\frac{\mu F}{1+\mu F/V_{sat}}
\end{equation}
where $\mu$ is the low-field mobility and $V_{sat}$ is the saturation velocity. A reference field $F_{ref}$, at which velocity shows saturation, can be defined as $F_{ref}=V_{sat}/\mu$. For the armchair direction, we find $F_{ref}=1.7$ V/$\mu$m and 3.3 V/$\mu$m for $n=10^{12}$ cm$^{-2}$ and $n=10^{13}$ cm$^{-2}$, correspondingly, and $F_{ref}=6.3$ V/$\mu$m and 14 V/$\mu$m for the zig-zag direction, respectively. Such high value sof $F_{ref}$ imply an absence of the velocity saturation, except for $n=10^{12}$ cm$^{-2}$ case in the armchair direction, when $V_{sat}=1.9\times10^{7}$ cm/s from the fit to Eq.~(\ref{eq:vsat}).

Nevertheless, the Joule losses lead to the temperature rise as $T=T_{amb}+jF/r$, where $T_{amb}=300$ K is an ambient temperature, $j=enV$ is the electric current, and $r$ is the thermal conductance. The upper bound for $r$, for a monolayer device on an insulating substrate, corresponds to the zero contact thermal resistance between the black phosphorus and the substrate. Such that $r$ is determined by the thermal conductivity and thickness of an insulating substrate. For example, for an oxide thickness with $h=300$ nm of SiO$_2$ substrate with thermal conductivity of $\kappa=1.4$ W/(mK) Ref.~\cite{SiO2thermal}, a maximum value of $r=\kappa/h\approx 0.47$ kW/(K cm$^2$) is expected.
Calculations with the self-heating effects included lead to the velocity saturation at much smaller field of $F_{ref}=0.6$ V/$\mu$m and saturated velocity of $V_{sat}=5.6\times 10^6$ cm/s, as shown in Fig.~\ref{fig:saturation}a. We also show in Fig.~\ref{fig:saturation}a an isothermal calculations at the burning temperature of black phosphorous few tens of nm thick sample~\cite{engel2014photodetector}. However, in the high bias experiments on few tens of nm samples in Ref.~\cite{engel2014photodetector} and \cite{Ahmed2016bp_burn} current was almost linear up to the break down voltage. This may imply an important  role of the electron-hole pair excitation across the small bandgap of $E_g\approx0.3$ eV in multilayers  at high temperatures and large electric fields. Such that increased $n$ would overwhelm a reduction of $V$ due to the saturation.

\section{Conclusions}\label{Sec:concl}

In conclusion, using a tight-binding model with parameters benchmarked by the DFT calculations, we have predicted phonon limited transport of black phosphorous as a function of temperature, carrier density, and electric field. The low-field mobility is highly anisotropic with values along the armchair direction about eight times  higher than that of the zig-zag direction. This difference stems from the high ratio of the effective masses and almost isotropic scattering rate. At room temperatures, optical phonons dominate the scattering leading to decreasing mobilities with increasing carrier densities. While transport data on monolayers are scare for a direct comparison, the available low bias measurements on black phosphorous multilayer samples are consistent with our predictions. We hope our work would stimulate further experiments on monolayer samples, which requires inert gas environments to prevent oxidation~\cite{Koenig2014bp_fet,Favron2015bp_oxidation,Edmonds2015bp_oxidation,Liu2015bp_oxidation}.

\appendix

\section{\\DFT derived parameters for the electron-phonon scattering}\label{Sec:DFT}

Using WIEN2k density functional theory (DFT) code~\cite{Blaha2014} within the PBE functional~\cite{Perdew1996:PRL}, for a monolayer black phosphorous we obtained the internal parameter $v=0.061$ and the equilibrium lattice constants of $a=4.63$ \AA~ and $b=3.31$ \AA~ along the armchair ($x$) and the zig-zag ($y$) directions, respectively. The layer separation in the  $z$ direction was set at 1.5 inter layer distance, i.e. $c=17.4$ \AA. (For the bulk black phosphorus we obtain values $v=0.110$ and $c=11.12$ \AA.)
The bandgap $E_g$ of 0.9 eV for a monolayer black phosphorus is consistent with the previous theoretical reports~\cite{Liu2014Egbp,Qiao2014,peng2014strain,Li2014Egbp_starin,tran2014excitons}.

\begin{figure}[h]
\centering
\includegraphics*[width=9.0cm]{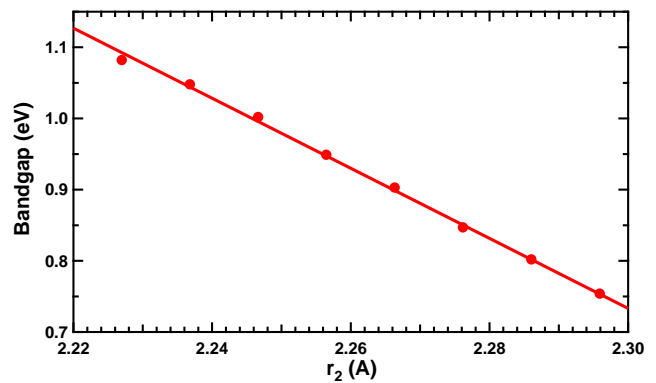}
\caption{(Color online) Monolayer black phosphorus DFT bandgap dependence on strain applied in the perpendicular to the plane direction (001). This strain direction keeps the in-plane P-P bondlength $r_1$ constant. The solid line corresponds to the best fit to Eq.~(\protect{\ref{eq:Eg_zz}}) with $g_2=2.46$ eV/\AA.
}\label{fig:Egstrain_zz}
\end{figure}

\begin{figure}[h]
\centering
\includegraphics*[width=9.0cm]{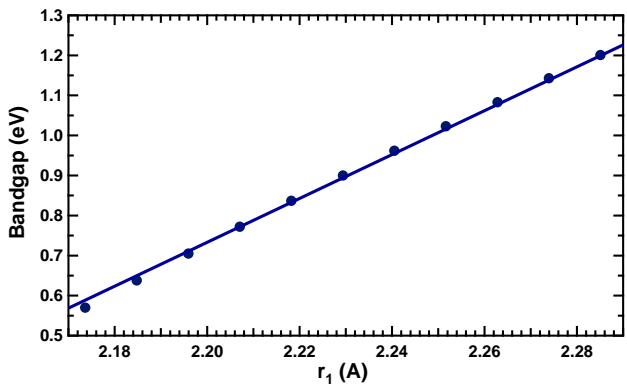}
\caption{(Color online) Monolayer black phosphorus DFT bandgap dependence on strain applied in the in-plane direction (110). This strain direction  keeps the in-plane bondangle constant.  The solid line corresponds to the best fit to Eq.~(\protect{\ref{eq:Eg_xy}}) with $g_1=1.53$ eV/\AA~ for a fixed  $g_2=2.46$ eV/\AA.
}\label{fig:Egstrain_xy}
\end{figure}

As we apply an out-of-plane strain in (001) direction, the out-of-plane bondlength $r_2$ increases, while the in-plane $r_1$ bondlength is constant. As a result, the bandgap reduces, as shown in Fig.~(\ref{fig:Egstrain_zz}). For the in-plane strain along the (110) direction, the in-plane bondangle is constant and the bandgap increases, as shown in Fig.~(\ref{fig:Egstrain_xy}).  Using a relationship between the bandgap and the hoping integrals from Eq.~(1) in the main text, i.e. $E_g=2t_2-4\vert t_1\vert$,  we can obtain:
\begin{eqnarray}
\frac{d E_g}{d r_{2}}&=&-2g_2 \ \ \ {\rm  \ \ \ \ \ \ \ \ \ \ \            for \ strain \ along \ (001)} \label{eq:Eg_zz} \\
\frac{d E_g}{d r_{1}}&=&4g_1-2g_2\frac{d r_2}{d r_1} \ \ \ {\rm for \ strain \ along \ (110) } \label{eq:Eg_xy}
\end{eqnarray}
where $dr_2/dr_1\approx 0.13$ is due to the geometrical factor, i.e. the out-of-plane P-P bond is not perpendicular to the plane.
The best fits of the bandgap change with an applied strain to Eq.~(\ref{eq:Eg_zz}) and (\ref{eq:Eg_xy})  result in the values of $g_1=1.53$ eV/\AA~ and $g_2=2.46$ eV/\AA.

\section{\\Relaxation time approximation for an anisotropic material}\label{Sec:RTA}

We are looking for a solution of the Boltzmann transport equation (BTE) for a carrier distribution function $f_k$ in an external electric field, without loss of generality, along the $x$ direction:
\begin{equation}
e F_x \frac{\partial f_k}{\partial p_x}=I_c=-\sum_{\vec{k'}} \Big[S_{kk'}f_{k}(1-f_{k'}) - S_{k'k} f_{k'} (1-f_{k})\Big]
\end{equation}
where $I_c$ is a collision integral, $S_{kk'}$ is an electron-phonon scattering rate, and $p_x=\hbar k_x$ is carrier momentum.

The distribution function can be written in the form
\begin{equation}
f_k = f_{0k}+f_{1k}
\end{equation}
where $f_{0k}$ is the equilibrium Fermi Dirac distribution. We assume that $f_{1k}$ is small and linear in $F_x$. In the collision integral, the zeroth order terms in $F_x$ sum to zero and the second order terms in field strength are discarded, such that:
\begin{equation}\label{eq_Ic}
\begin{split}
I_c = &- \sum_{\vec{k'}} f_{1k} \Big[ S_{kk'} (1-f_{0k'}) + S_{k'k} f_{0k'} \Big]
\nonumber \\
&+\sum_{\vec{k'}}f_{1k'} \Big[ S_{kk'} f_{0k} + S_{k'k} (1-f_{0k}) \Big] \\
\end{split}
\end{equation}
Using the detailed balance relationship $S_{kk'}f_{0k}(1-f_{0k'}) = S_{k'k} f_{0k'} (1-f_{0k})$,  Eq.~(\ref{eq_Ic}) transforms to
\begin{equation}
\begin{split}
I_c &= -\sum_{\vec{k'}} S_{kk'} \Big( f_{1k} \frac{1-f_{0k'}}{1-f_{0k}}- f_{1k'} \frac{f_{0k}}{f_{0k'}} \Big) \\
\end{split}
\end{equation}
Using an ansatz for the distribution function
\begin{equation}
f_{1k} = e F_x \frac{\partial E_k}{\partial p_x} \Big( -\frac{\partial f_{0k}}{\partial E_k} \Big) \tau( \vec{k})
\end{equation}
the collision integral takes the form:
\begin{equation}
\begin{split}
I_c &= - \frac{e F_x}{kT} \sum_{\vec{k'}} S_{kk'}f_{0k}(1-f_{0k'}) \Big[ v_{kx} \tau( \vec{k}) - v_{k'x} \tau( \vec{k'}) \Big]
\end{split}
\end{equation}
where $v_{kx}=\partial E_k/\partial p_x$ is the band velocity. The left hand side of the BTE, which is linear in $F_x$, is given by:
\begin{equation}
e F_x \frac{\partial E}{\partial p_x} \frac{\partial f_{0k}}{\partial E_k}= -\frac{e F_x}{kT} v_{kx} f_{0k}(1-f_{0k})
\end{equation}
such that the BTE equation reduces to:

\begin{equation}
\begin{split}
1 = \sum_{\vec{k'}} S_{kk'} \frac{1-f_{0k'}}{1-f_{0k}} \Big[ \tau(\vec{k}) - \frac{v_{k'x}}{v_{kx}} \tau(\vec{k}') \Big]
\end{split}
\end{equation}

If the scattering is nearly elastic and scattering time has weak angle dependence,
i.e. $\tau(\vec{k}')\approx\tau(\vec{k})$, we can arrive to the following expression for the relaxation rate and mobility:
\begin{equation}\label{eq_mob}
\begin{split}
&\frac{1}{\tau_x(\vec{k})} = \sum_{\vec{k'}} S_{kk'} \frac{1-f_{0k'}}{1-f_{0k}} \Big( 1 - \frac{v_{k'x}}{v_{kx}} \Big) \\
\mu_x &= \frac{ j}{enF_x} = \frac{\int d\vec{k} v_{kx} f_{1k} }{F_x \int d\vec{k} f_0} =
 \frac{e \int  d \vec{k}  \left(-\frac{\partial f_{0k}}{\partial E_k}\right) v_{kx}^2 \tau_{kx}}{ \int d\vec{k} f_{0k}}
\end{split}
\end{equation}
where $j$ is an electric current density and $n$ is the carrier density. Eq.~(\ref{eq_mob}) coincides with Eq.~(\ref{eq:scat})-(\ref{eq:low_mob}).

\end{document}